# Dynamics of Insect Paraintelligence:
## How a Mindless Colony of Ants Meaningfully Moves a Beetle

Eldar Knar[1]

Tengrion, Astana, Republic of Kazakhstan

eldarknar@gmail.com
https://orcid.org/0000-0002-7490-8375

**Abstract**

In this work, a new concept called *Vector Dissipation of Randomness* (VDR) is developed and formalized. It describes the mechanism by which complex multicomponent systems transition from chaos to order through the filtering of random directions, accumulation of information in the environment, and self-organization of agents. VDR explains how individual random strategies can evolve into collective goal-directed behavior, leading to the emergence of an ordered structure without centralized control.

To test the proposed model, a numerical simulation of the "ant–beetle" system was conducted, in which agents (ants) randomly choose movement directions, but through feedback mechanisms and filtering of weak strategies, they form a single coordinated vector of the beetle's movement.

VDR is a universal mechanism applicable to a wide range of self-organizing systems, including biological populations, decentralized technological networks, sociological processes, and artificial intelligence algorithms.

For the first time, an equation of the normalized emergence function in the processing of vector dissipation of randomness in the Ant–Beetle system has been formulated.

The concept of *paraintelligence* was introduced for the first time. Insect paraintelligence is interpreted as a rational functionality that is close to or equivalent to intelligent activity in the absence of reflexive consciousness and self-awareness.

**Keywords:** paraintelligence, paraconsciousness, vector dissipation of randomness, self-organization, randomness, order, phase transition, system emergence, collective behavior.

**Declarations and Statements:**
No conflicts of interest
This work was not funded
No competing or financial interests
All the data used in the work are in the public domain.
Generative AI (LLM or other) was not used in writing the article.
Ethics committee approval is not needed (without human or animal participation).

---

[1] Fellow of the Royal Asiatic Society of Great Britain and Ireland

# 1. Introduction

Static states are most often interpretable, and the orderliness of inanimate conditions is fairly well studied. However, the dynamics and order emerging from unintentional actions require scientific examination and discussion. A fundamental question arises here—how does unconscious supercollective behavior lead to a conscious and specific result through the mechanism of emergence?

That is, how do qualitatively new, meaningful, and functional structures, patterns, or solutions arise from numerous individual actions that initially lack conscious or purposeful character? There are many such examples in both living and nonliving natures, as well as in economics, society, social networks, and other societal domains.

The spontaneous or deterministic emergence of system-level behavior does not obviously follow from the behavior of the individual elements of the system. Elements do not aim—consciously or unconsciously—to achieve emergence; each element pursues its own local and individual goals. There is no unified plan.

However, gradually, through a series of iterations, simple individual actions begin to synchronize, merging into unconscious but effective collective behavior with the formation of stable patterns. When a temporally defined critical mass of such unconscious actions is reached, a phase transition occurs. Collective behavior becomes stable, comprehensible, and begins to be interpreted by external observers as something that is goal oriented and even conscious.

In complex systems, numerous elementary interactions at the microlevel lead to a phase transition, wherein behavioral structures or components qualitatively change upon reaching a critical threshold of interaction (e.g., percolation). This triggers a critical transformation in the system's behavior, ultimately forming a seemingly "meaningful" structure (e.g., rational behavior from irrational objects). The feedback mechanisms within the system provide a certain degree of stability to this process.

Once emergent behavior is formed, agents begin to perceive it, adjust their behavior on the basis of the new state, and it becomes reinforced—turning into a conscious and stable result. In general, randomness and unconsciousness at the level of individual elements create clearly defined, meaningful, and goal-oriented outcomes at the level of the whole system.

However, randomness in collective behavior does not simply transform into some kind of order. Randomness itself can sometimes produce a form of order—so to speak, purely by chance. In particular, randomness condenses meaning. Collective unconscious behavior does not merely "become conscious." It inevitably compresses into a structure, such as a salt crystal, a protein, or information. A meaningful result arises as the least costly, energetically optimal structure. In this sense, meaning is the final stage of randomness compression.

There is meaning in the compression of randomness. Suppose that the number of agents performing random actions increases—at some point, the information begins to structure itself, and a unified formula or behavioral algorithm with minor fluctuations emerges. That is, randomness naturally compresses into an ordered algorithm. Here, randomness is interpreted as a raw informational field, information as structured randomness, and a behavioral algorithm as compressed, optimized information.

Moreover, collective randomness has a "saturation threshold," after which the system begins to self-organize. When the number of agents exceeds a certain critical limit, the system "chooses" the most energy-efficient behavioral algorithm. The system always selects the simplest algorithm from all available options. The more agents there are, the stronger this effect becomes. This is precisely why collective unconsciousness always leads to a meaningful result—the system simply discards unnecessary noise.

When an unconscious system collapses into an algorithm of "conscious" behavior, randomness does not vanish completely—and localized fluctuations remain.

Let us assume that a colony (group) of ants discovers a beetle. Initially, their behavior is random—they simply swarm around and move chaotically near the beetle. Each individual has a set of behavioral vectors, e.g., 10 or even 100 random directions.

One of these random vectors accidentally leads to a positive result—for example, a push or movement in the direction of the anthill. This may be determined by the recognition of pheromone scent trails or orientation-based perception. The intensification of a familiar trail odor or the scent of the nest itself acts as a signal of a correct random action.

Moreover, an ant may also orient itself according to the number of other ants present. The closer it is to the anthill, the more ants it encounters. In fact, ants constantly run along the path to and from the beetle to create a perceptual anchor through the "presence effect." Gradually, the vector of correct action or behavior is reinforced and stabilized, whereas other random vectors dissipate. As a result, a collective formula emerges for a unified, random—yet ultimately correct—behavioral vector.

This seemingly trivial mechanism of self-organization explains how a multitude of random behavioral vectors collapse into a single formula of collective action. Thus, random collective behavior inevitably transforms into a meaningful emergent result.

In essence, the ant–beetle system functions as a natural computational algorithm that filters randomness, retaining only the optimal behavioral vector. All vectors start with equal probability. The first successful random event produces a positive signal. The ants begin aligning with this vector, as it amplifies the feedback loop. The more iterations that occur, the more efficiently nonoptimal random vectors are filtered out. This filtering effect is strengthened by the presence effect (as ants run back and forth, reinforcing spatial cues) and by the amplification of signals through pheromone concentration and the increasing number of agents. In the final iterations, only a single emergent vector remains—around which only minor fluctuations may persist.

Thus, "purposeful" behavior can be interpreted as the result of natural randomness filtering.

However, simple copying of another's behavior is not meaningful in itself. An ant imitates another only when the latter's behavior aligns with its internal predispositions. For example, another ant may randomly push the beetle, and by chance, the beetle moves in the direction of the nest or along a familiar trail—or towards a cluster of other ants. This action then becomes an amplifier and a trigger for copying due to enhancement of the pheromone field or the activation of spatial orientation cues. Therefore, the ant is not merely copying—it calibrates its behavior, using the actions of others as confirmation of the correct vector.

In this sense, the copying process is not just social behavior but also a mechanism for calibrating the agent's own model of the world.

Emergence is generated not only through mechanical adjustment, imitation, and positive feedback. It can also occur at the individual level. That is, each ant independently arrives at the correct algorithm. Together, they form a unified behavioral algorithm. In other words, each ant individually computes the correct response—and then all of their individual solutions synchronize. This is not simply a collective filtering of randomness but a process of individual adaptation by each agent—a type of individual biophysical learning.

After each ant has learned independently, their individual algorithms begin to converge. This occurs because all ants operate within the same local signal environment. Especially in the ant–beetle system, agents inevitably interact with one another, reinforcing the effect. The final algorithm is fixed in the collective probability distribution and is interpreted as a statistical sum—a collective solution.

Thus, the overall algorithm is not formed top-down but rather bottom-up. Individual agents independently discover the same solution because they share a common "solution space." The environment itself acts as an "invisible teacher" that simultaneously educates all individuals in the same way. This is where the fundamental effect of informational compression becomes evident.

Therefore, random behavior is not merely filtered—it evolves individually within each agent. However, since the "solution space" is identical for all, each agent arrives at the same "correct" emergent strategy. This is more than feedback—it is the collective result of individual learning. Each ant "computes" the correct algorithm independently rather than simply imitating others. What appears to be "collective intelligence" can in fact be interpreted as the result of massive independent learning in a shared environment. In this case, collective intelligence functions most efficiently in the absence of centralized control. In the ant–beetle system, there are no leaders and no subordinates.

Overall, this is a hybrid process in which individual behavior, combined with filtering, behavioral imitation, and a range of environmental factors, produces emergence. Irrational (nonoptimal) random vectors are compressed out, leaving behind a rational behavioral structure in the form of "meaningful" action patterns.

This interpretation—the compression of all possible random behavioral vectors into a single, stable emergent dynamic—is what we define as *Vector Dissipation of Randomness (VDR)*.

Vector dissipation of randomness is understood as a process in which a multitude of random behavioral vectors, inherent to individual agents within a complex system, collapse into a stable and meaningful structure through mechanisms of individual learning, resonance-based imitation, and the filtering of weak signals at an "unconscious" level. VDR can explain how systems composed of many independent, unconscious agents are capable of generating global, "conscious" behavioral algorithms—without any centralized control.

## 2. Literature Review

Functional neuroscience (Frank & Kronauer, 2024) and concepts of self-organization (Sumpter, 2006) are actively employed to understand collective behavior in animals (Sumpter, 2006). In collective animal behavior, group-level patterns emerge from individual-level interactions among members (Mizumoto & Reid, 2024). In particular, complex models arise in social insects with mechanisms of task allocation (Beshers & Fewell, 2001). For analysing cooperative behavior and collective intelligence in biological groups, qualitative models of agent-based group dynamics with informational leadership are especially relevant (Fu et al., 2024). Similarly, quantitative models of numerical simulations in biosocial networks (Naito et al., 2024), which are also utilized in our study, have proven to be increasingly significant.

Despite the fundamental similarities between models of collective and herding movement (Vicsek & Zafeiris, 2010), there are serious challenges in capturing some of the more specific and important aspects. In particular, we argue that such models may also pertain to manifestations of rational activity, which is often interpreted as collective intelligence (McMillen & Levin, 2024) or swarm intelligence (Garnier et al., 2007). Of course, this is not "intelligence" in the full sense of the word; as such, rational behavior excludes consciousness—understood by some researchers as integrated information with a set of informational relations generated within a system (Tononi, 2008).

At the same time, the presence of collective memory in animal groups (Couzin et al., 2002) may serve as a foundation for prototyping analogues or parallels with intelligent—or even conscious—activity in social animals and insects (in certain manifestations).

Collective intelligence facilitates resource allocation through frequency-dependent learning (Ogino & Farine, 2024). That is, group-living animals distribute resources more efficiently than solitary animals do. In particular, the efficiency of such a distribution is well illustrated by nest-site selection algorithms, which use emigrating ant colonies (Temnothorax) to reach a consensus on a new nest site (Zhao et al., 2020), and by the effective orientation of ants in unstructured environments (Gelblum et al., 2020).

Moreover, collective behavior often implies the potential or actual emergence of emergent phenomena (Cucker & Smale, 2007), which is a key focus of our study in the context of the ant–beetle dynamic system. In particular, ants exhibit an emergent collective sensory response threshold, which depends on group size and is driven by social feedback mechanisms among individuals (Gal & Kronauer, 2021).

In general, mass effects and aggregation in systems of simple individuals can lead to efficient and reliable solutions to environmental challenges through self-organization, adaptation, and other parameters of interaction and connectivity (Priya et al., 2024). Adaptability, in particular, is one of the dominant strategies for the emergence of collective intelligence (Falandays et al., 2023).

Indeed, group-level animal behavior can qualitatively overcome the limitations of individual behavior in solving the cognitive tasks faced by biological communities and populations (Krause et al., 2010). Conversely, individual-level diversity in form and behavior creates the necessary conditions for collective emergence—particularly in ant colonies.

In our work, we also conduct a comparative analysis of models of collective intelligence, self-organization, and emergence. Specifically, we consider ACO—ant colony optimization theory (Dorigo, 2005), PSO—particle swarm optimization (Assareh, 2010), SOM—self-organizing maps (Lawrence, 1997), and the Ising model (Onsager, 1944).

3. **Methodology**

The model interpretation was developed via the Python programming environment. The program was implemented, and scenario simulations were run in the interactive Jupyter Notebook environment with predefined technical parameters (Table 1).

*Table 1.* Technical parameters of the numerical model structure

| Parameter | Meaning | Description |
|---|---|---|
| Time step | $\Delta t = 0.1$ | Discrete time of one iteration of simulation |
| Total number of iterations | $T_{max} = 5000T$ | The time during which the system's movement is analysed |
| Number of agents | $N \in [100, 10000]$ | Range of number of ants participating in the experiment |
| Critical number of agents | $N_c \approx 5000$ | Approximate value at which the system enters an ordered state |
| Adaptation coefficient | $\lambda = 0.1$ | Determines how quickly ants adapt to successful strategies |
| Weak Strategies Extinction Coefficient | $\gamma = 0.05$ | The higher, the faster the system gets rid of chaotic directions |
| Filtering coefficient of random strategies | $\delta = 0.2$ | Shows how quickly ants collectively arrive at a dominant direction |

To simulate probabilistic changes in agent strategies, the Monte Carlo method was employed (Table 2). Agent-based modelling (ABM) was used to simulate the behavior of individual ants, their interactions, and the filtering of random directions (Table 2).

*Table 2.* Dominant variables of the model scheme

| Variable | Designation | Description |
|---|---|---|
| Number of ants | $N$ | The number of agents in the system involved in the movement of the beetle |
| The position of the beetle | $x(t)$ | The current coordinate of the beetle on the axis of movement |
| Beetle speed | $v(t)$ | Derivative of $x(t)$, the speed of movement of an object |
| Number of directions of movement of an ant | $V$ | Number of possible vectors of direction of movement (e.g. 8 directions) |
| Probability of choosing direction v by ant i | $P(v, t)$ | The probability with which an agent chooses a particular motion vector at time t |
| General direction of movement of the system | $S(N, t)$ | Measure of orderliness of the system at time t |
| Force applied by an ant | $F$ | The force with which the i-th ant pulls the beetle |
| Total force applied to the beetle | $F(t)$ | The sum of all F directed at an object |
| The mass of the beetle | $M$ | The mass of an object that affects its inertia |

| Pheromone trail at point x | $\varphi(x,t)$ | Concentration of pheromones in the environment at time t |
|---|---|---|
| Agents adaptation coefficient | $\lambda(t)$ | Shows how quickly the agent copies successful strategies |
| Forgetting rate of weak strategies | $\gamma(t)$ | Determines the speed of "erasing" unsuccessful directions of movement |
| Critical number of agents | $N_c$ | The number of ants at which the beetle begins to move steadily |
| Random Strategy Filtering Threshold | $\delta$ | Determines how quickly the system gets rid of ineffective directions |
| Pheromone accumulation constant | $\rho$ | A coefficient describing the rate of evaporation and accumulation of traces |

The model primarily accounts for the filtering of random strategies, information accumulation in the environment, and networked interactions among the ant agents.

## 4. Results

The emergent behavior function E(t) in the process of Vector Dissipation of Randomness is interpreted as the transformation of the ant–beetle system from chaotic wandering to coordinated movement of the beetle through agent adaptation, learning, information accumulation, and the filtering of random strategies. In our case, the dominant characteristic of this process is the change in correlation between agents over time.

Accordingly, the formalized model of vector dissipation of randomness must be as follows:

normalized (0≤E(t)≤10,

invariant across different systems,

coordinated (dependent on the number and degree of agent coordination),

integrated in the context of dynamic changes in ant-agent characteristics.

On this basis, we define the emergent function as a multiplicative composition of three parameters:

$$E(t) \propto C(t) \cdot S(t) \cdot A(t)$$

where

E(t) – dynamic emergent behavior in the ant–beetle system,

C(t) – directional correlation (orientation) among ant agents,

S(t) – accumulated environmental information level,

A(t) – adaptivity level of the ant agents.

This representation is relatively simple but logical and authentic for an approximate model.

The directional correlation is defined as a cosine-based function reflecting how closely the ants align their movement:

$$C(t) = \frac{1}{N} \sum_{t=1}^{N} \cos(Q_i + Q)^2$$

where
$Q_i$ – Movement angle of the i-th ant,
$Q$ – Average movement angle of the group.

If all the ants move in the same direction, then $C(t) \to 1$; if the movements are chaotic, $C(t) \approx 0$. In reality, ants generally do not push prey; rather, they drag it toward the nest. However, some individuals do push, so the key is not the orientation of the ants themselves but the direction of movement of the ant–beetle system.

The accumulated information level incorporates pheromones, visual perception, vibrations, and feedback. The higher these values are, the greater the amount of stored information. It is modelled with a logistic sigmoid function:

$$S(t) = 1/(1 + exp(-\beta(N - N_c)))$$

where
$N_c$ - critical number of agents required for emergence,
$\beta$ - system sensitivity to information accumulation.

If $N \ll N_c$, then $S(t) \approx 0$; if $N > N_c$, then $S(t) \approx 1$.

The adaptivity of the ant agents, which is based on individual learning and environmental interaction, is modelled as follows:

$$A(t) = 1 - exp(-\gamma \cdot t)$$

where
$\gamma$ – rate of agent adaptation.

Initially, $A(t) \approx 0$ (no adaptation), and over time, $A(t) \to 1$.

To capture the critical point of phase transition, we introduce the transition speed parameter α\alphaα. Thus, the full emergent function becomes:

$$E(t) = \frac{C(t) \cdot S(t) \cdot A(t)}{1 + exp(-\alpha(t - t_c))}$$

where
α - speed of the phase transition,
$t_c$ - critical time point of emergence.

The term 1+exp(−α(t−$t_c$)) acts as a normalizing sigmoid function, describing a smooth transition from chaotic behavior to ordered, goal-directed movement.

Therefore, the emergent function is characterized by a smooth phase transition and remains within the range 0≤E(t)≤10.

We refer to this as the normalized emergence function (NEF).

Hence, the NEF captures directional correlations, environmental information accumulation, agent adaptivity, and the critical transition zone (Table 3).

*Table 3.* Interpretation of the phase transition in the ant–beetle system

| Phase | Processing | Dynamics E(t) |
| --- | --- | --- |
| Before emergence (t< $t_c$) | Ants act randomly, information accumulation is minimal | E(t)≈0 |
| Range emergence (t≈$t_c$) | The ants begin to move synchronously, information is actively accumulated | E(t) increases sharply |
| After emergence (t> $t_c$) | The ants move the beetle in a coordinated manner, the system stabilizes | E(t)≈1 |

In principle, the normalized emergence function may be applied to a wide range of self-organizing systems—from swarm intelligence to neural networks and biological populations.

In simplified form, the NEF can be expressed as:

$$E(t) = \frac{t}{(t + T_C)} \cdot \frac{N}{(N + N_C)} \cdot (1 - e^{-t/T_A})$$

Here, we rotate the formula from angular correlation to a simpler time-based dependency. Instead of a logistic function, a smooth S-shaped function is used, and adaptivity is modelled with a basic exponential term.

This simplified NEF still preserves the core effects and processes (correlation, information accumulation, adaptivity).

*Table 4.* Phase transition interpretation in the simplified ant–beetle system

| Phase | How does E(t) change? | What's happening? |
|---|---|---|
| Chaos (t≈0) | E(t)≈0 | Ants act randomly, there is no accumulated information |
| Emergence (t≈$T_C$) | E(t) increases sharply | The system begins to self-organize, filtering random directions |
| Orderliness (t ≫ $T_C$) | E(t)≈1 | The ants move the beetle in one direction, the system is stabilized |

The ant–beetle system does not shift into ordered movement instantly but rather undergoes a gradual self-organization process.

*Figure 1.* Temporal dynamics of system order relative to the number of ant agents

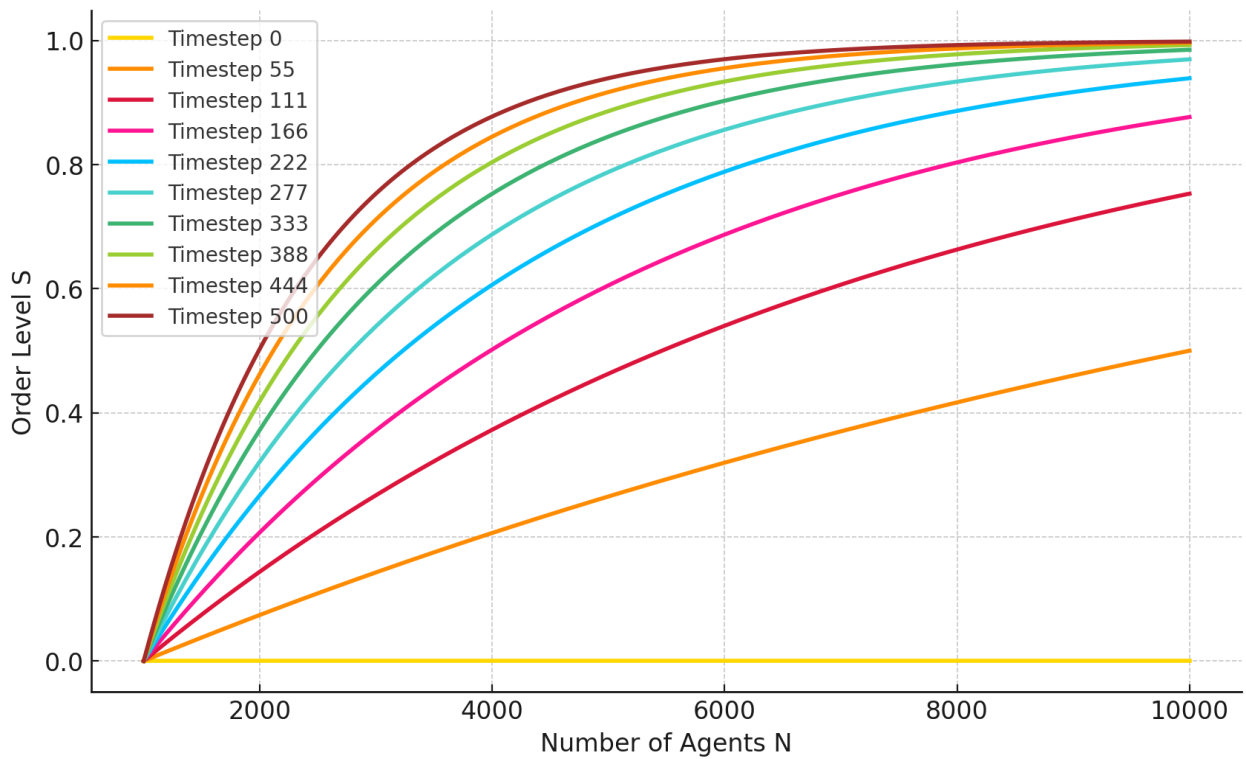

The degree of order, as a process of information accumulation and structural evolution, develops over time depending on the number of ant agents—until the point of maximum "meaningfulness" of the agents' collective behavior is reached.

In the initial time steps (0–166), ant actions are chaotic, and the level of coordination in collective effort grows slowly. In the intermediate time steps (222–

388), a process of intensive filtering of random directions begins, resulting in a rapid increase in the degree of order. In the later time steps (444–500), the system reaches a stable level of organization: the ants move the beetle in a correct and target-directed direction.

Overall, the ordering process takes the form of an S-shaped curve (Figure 2).

*Figure 2.* Integral interpretation of a smooth phase transition

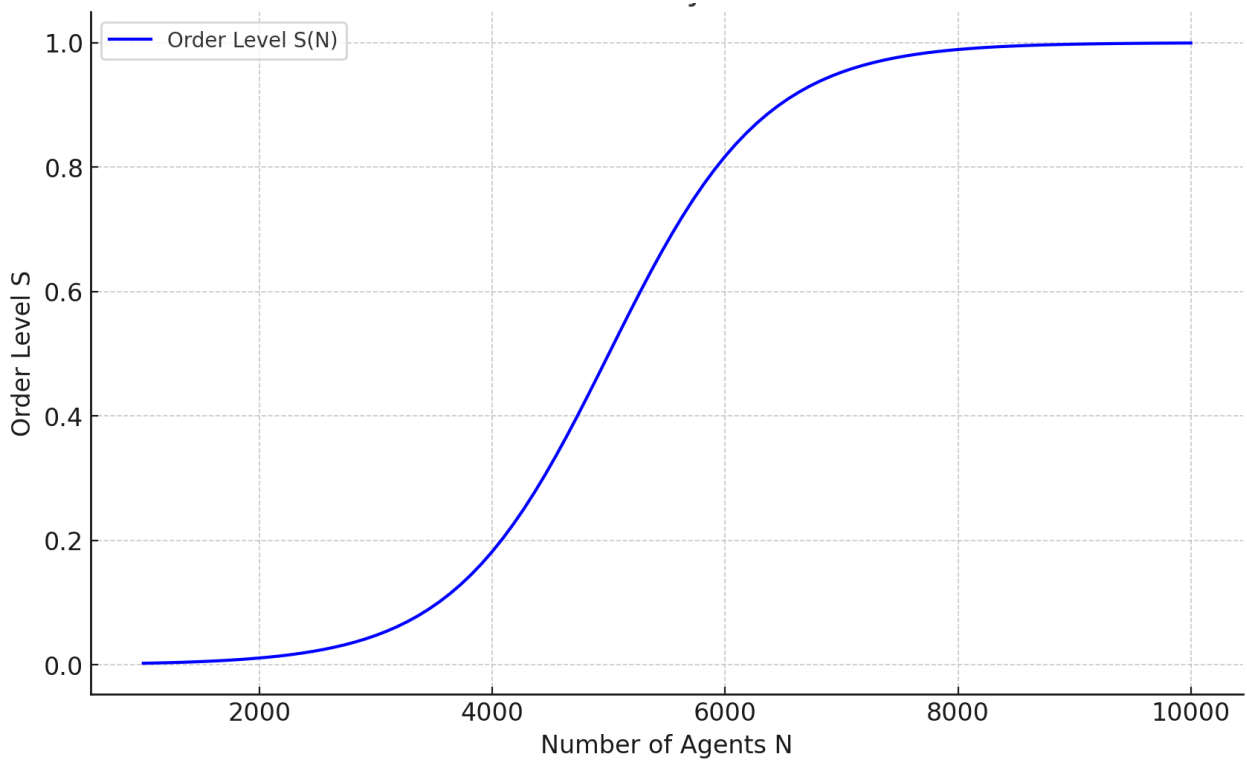

The dynamics of the ant–beetle system over time (0–500) reflect the cumulative process of gradual self-organization, which we interpret through three main stages of the system's self-organization:

*Table 5.* Three stages of self-organization in the ant–beetle system

| Phase | Description |
| --- | --- |
| Chaos phase (small N) | Agents act independently, random strategies are not filtered |
| Transition phase (average N) | Filtering of random strategies begins, the system self-organizes |
| Phase order (large N) | The system reaches a stable state, chaos disappears |

Thus, random directions do not disappear immediately. Their filtering occurs gradually until a single dominant vector of "meaningful" movement remains. This process is consistent with percolation models, where order arises only after a critical threshold of connectivity is reached.

The critical agent threshold N may vary (in our model, it was 5000), depending on the physical parameters and relationships in the system. In general, however, self-organization does not begin because of the presence of all agents at once but rather because of increased collective interaction and the achievement of a critical mass necessary for a phase transition.

In the context of phase transition, the critical parameters of emergence depend on the variability in agent adaptation (Figure 3).

*Figure 3.* Temporality of critical parameters of emergence

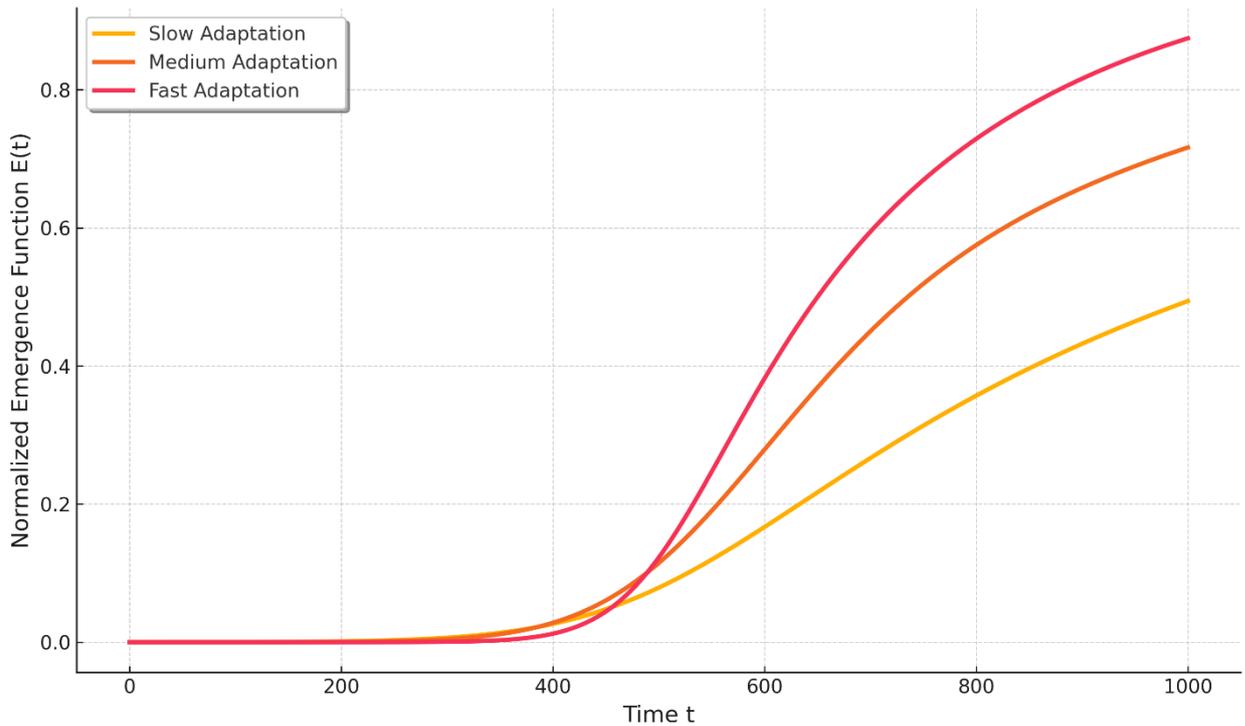

With faster adaptation, the phase transition occurs earlier (Table 6).

*Table 6.* Critical points of the phase transition

| Scenario | Time of maximum growth (critical point) | Maximum rate of change dE/dt |
| --- | --- | --- |
| Slow adaptation | 638.2 | 0.0010 |
| Average adaptation | 603.0 | 0.0018 |
| Fast adaptation | 562.8 | 0.0027 |

Accordingly, the greater the adaptivity is, the higher the rate of change dE/dt, and the sharper the increase in order (Figure 4).

*Figure 4.* Rate of change of the emergence function dE/dt

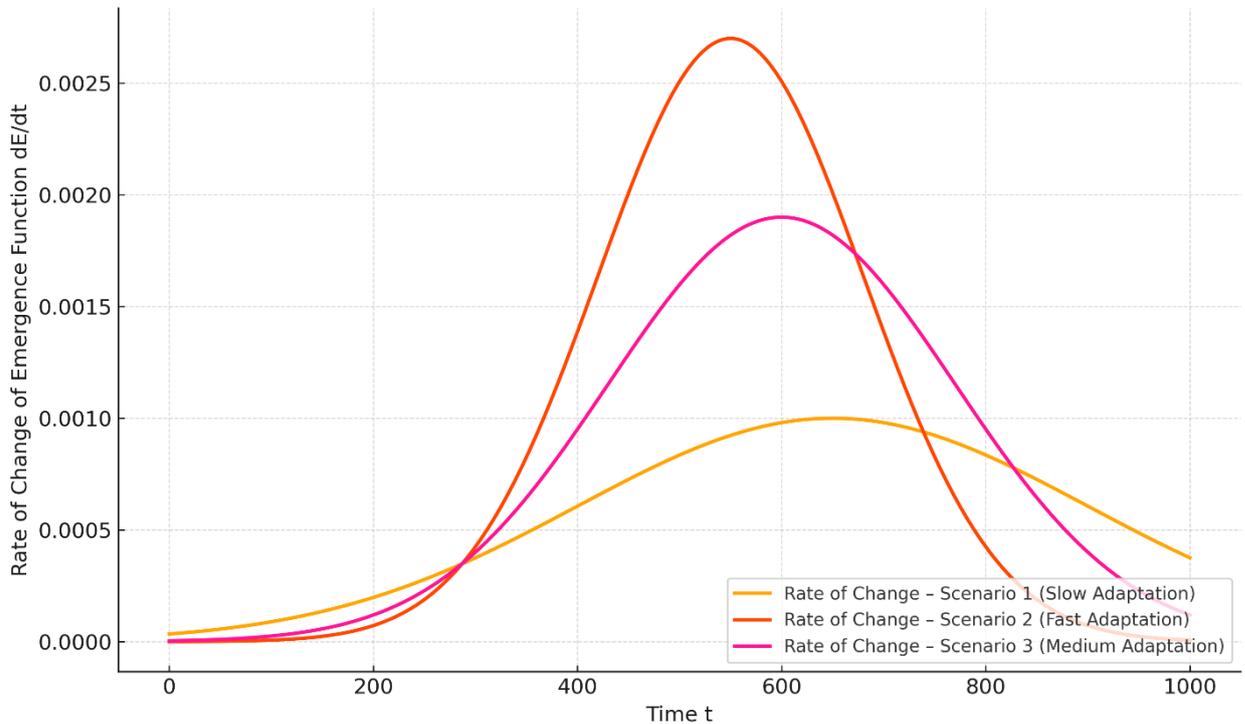

With slower adaptation, the phase transition is delayed, and the increase in coordination is smoother.

In summary, the higher the speed of adaptation is, the greater the overall emergence of the ant–beetle system. Rapid adaptation of ant agents leads to a sharp but effective transition to an ordered state. Slower adaptation results in a gradual increase in order, but the overall level of emergence becomes somewhat lower.

The phase transition does not occur instantly, as in classical physical models, but is spread over a time interval (Figure 5). With faster adaptation, the transition begins earlier. With slower adaptation, the system remains longer in a chaotic state.

Moderate adaptation provides a balanced dynamic.

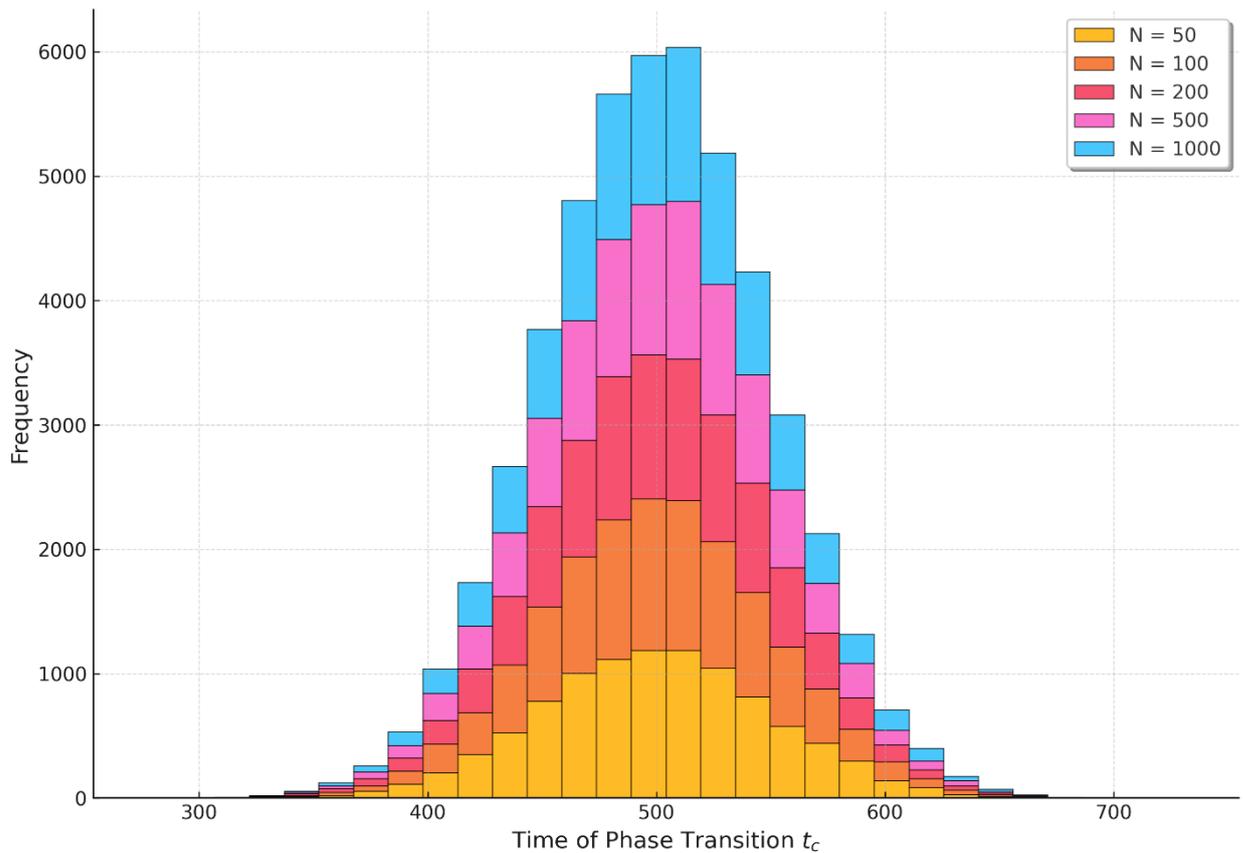

*Figure 5*. Histogram of the phase transition distribution

For a smaller number of agents, the phase transition occurs in a narrower range since the system organizes more quickly. At a larger number, the range widens, since more time is needed to coordinate all active participants of the ant–beetle system.

The influences of random fluctuations, the information dissipation rate, resistance to learning, and the initial distributions of the ant–agent directions all have similar and proportional dependencies.

The influence of external fields (pheromone gradients, vibrations, visual perception, etc.) also modulates the degree of organization in ant-agent behavior (Figure 6). In particular, ants perceive other ants not so much visually but rather through the "auditory" perception of vibrations. Naturally, the intensity of vibration and sensitivity to it depend on the number of ants in the system. This is an agent-based function, as are pheromones, visual cues, astronomical orientation, and so on.

This can be analogized to the behavior of nematic liquid crystals in a magnetic field. Without an external field, nematics are disordered. As the magnetic field increases, they begin to align in one direction. The same occurs with the ant colony: the pheromonal field, nest coordinates, and ant clustering act as analogues of the magnetic field. Under the influence of this "field," the ants begin to align collectively—unconsciously pushing the beetle in the correct direction. The stronger this field becomes, the more chaos is converted into order.

Without an external field, emergence is interpreted as a product of internal adaptation and filtering only. When the external integrated field is intensified, the phase transition occurs faster, and the final degree of order is higher.

*Figure 6.* Effects of External Fields on Emergence

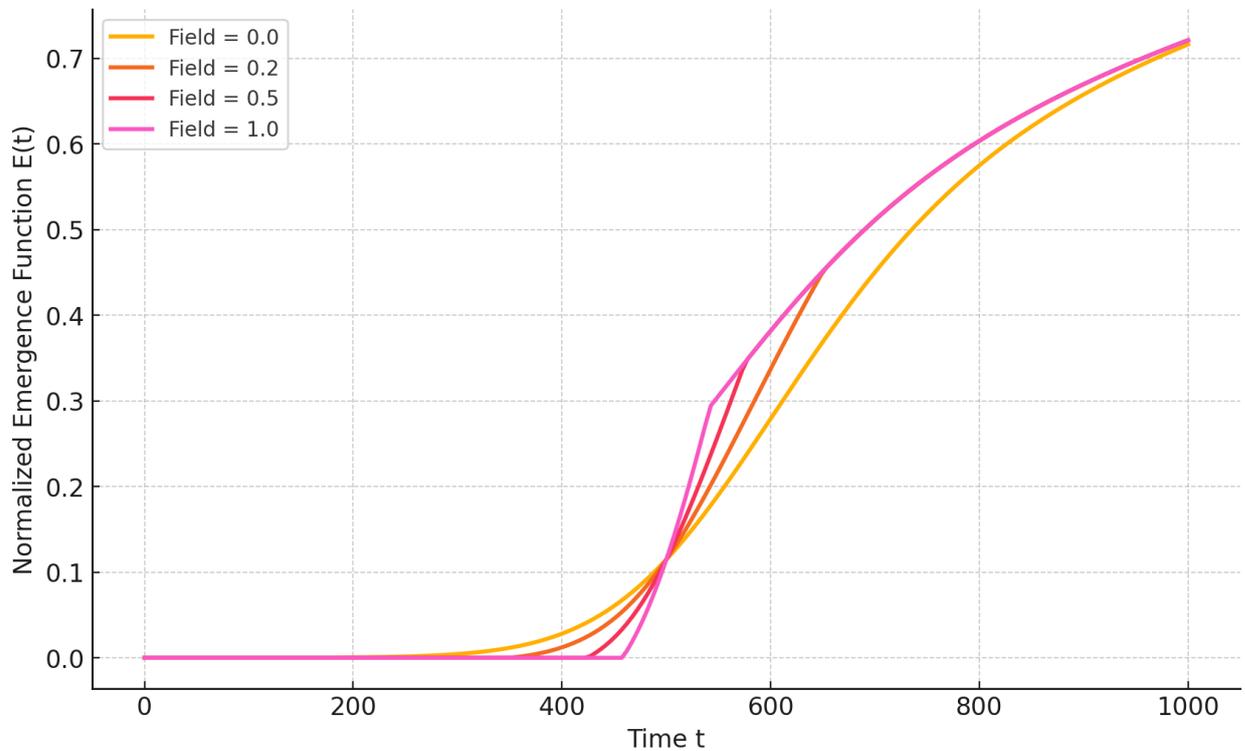

We interpret the cumulative results of how different parameters influence emergence in the context of vector dissipation of randomness in an integrated table (Table 7).

*Table 7.* Influence of key parameters on the behavior of the emergence function

| Factor | Minimal effect | Maximum effect | Conclusion |
| --- | --- | --- | --- |
| Adaptation speed | $\gamma = 0.002$ | $\gamma = 0.02$ | Increasing $\gamma$ sharply accelerates the phase transition |
| Information sensitivity | $\beta = 0.0005$ | $\beta = 0.005$ | The higher the $\beta$, the faster the environment is saturated with information |
| The power of fluctuations | $\sigma = 0.2$ | $\sigma = 0.01$ | High fluctuations smear the phase transition |
| Initial noise directions | Noise = 1.0 | Noise = 0.0 | The less the initial noise, the faster the ordering |
| External field strength | Field = 0.0 | Field = 1.0 | The external field dramatically accelerates and enhances emergence |
| Resistance of agents | ζ = 0.9 | ζ = 0.0 | Resistance strongly inhibits self-organization |

Here, we present six functional variants of the normalized emergence function E(t) with different combinations of parameters (Figure 7).

*Figure 7.* Combined Visualization: Effects of Parameters on Emergence

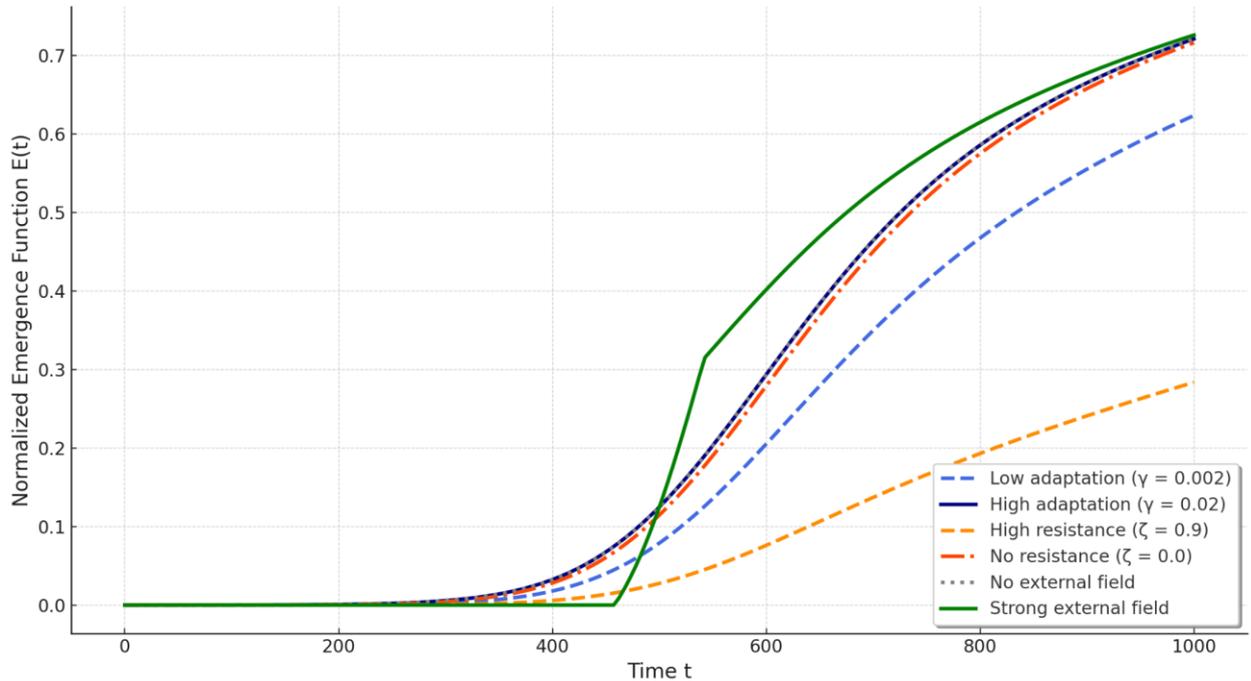

The six dynamic parameters represent the dominant behavioral aspects of agents in the vector dissipation of randomness model (Table 8):

*Table 8.* Parametric aspects of ant-agent behavior

| Parameter | Meaning | Interpretation |
|---|---|---|
| Adaptation (γ) | 0.002 | Slow adaptation |
| Adaptation (γ) | 0.02 | Fast adaptation |
| Resistance (ζ) | 0.9 | High cognitive inertia |
| Resistance (ζ) | 0.0 | Complete lack of inertia |
| External field (φ) | 0.0 | No landmark (the anthill is not felt) |
| External field (φ) | 1.0 | Strong reference point (direct influence) |

In this model, rapid adaptation is interpreted as a key factor accelerating self-organization. At low adaptation levels (γ=0.002), the curve grows smoothly with a prolonged latent period, and the ant–beetle system remains in a state of randomness for a long time. At high adaptation levels (γ=0.02), we observe a sharp increase in

E(t) with an early phase transition (approximately t=500). In this case, the ant agents rapidly adjust and develop a common strategy.

The resistance to learning suppresses the phase transition, making the system inert. With high resistance (ζ=0.9), the curve barely increases. Even in the presence of information and correlation, the ant agents do not adopt new rational strategies. With no resistance (ζ=0.0), the curve demonstrates steady growth and reaches a high level of order and rational behavior.

The external field acts as a trigger for a collective phase transition into an emergent state. In the absence of a field (ϕ=0.0), emergence develops only through internal interactions, with delay and incompleteness. In a strong external field (ϕ=1.0), such as high levels of motivating pheromones, the curve rises sharply, especially after t=500. This illustrates the role of oriented external fields in enhancing coordinated rational actions.

Thus, emergence in the vector dissipation of randomness (VDR) model results from the nonlinear superposition of internal and external factors (Figure 8). The internal parameters (adaptation and resistance) determine the system's readiness for qualitative transformation. External factors (information density, spatial cues, and external fields) shape the quantitative trajectory of the transition from randomness to order.

*Figure 8.* Final Emergence vs. Adaptation Speed and Resistance

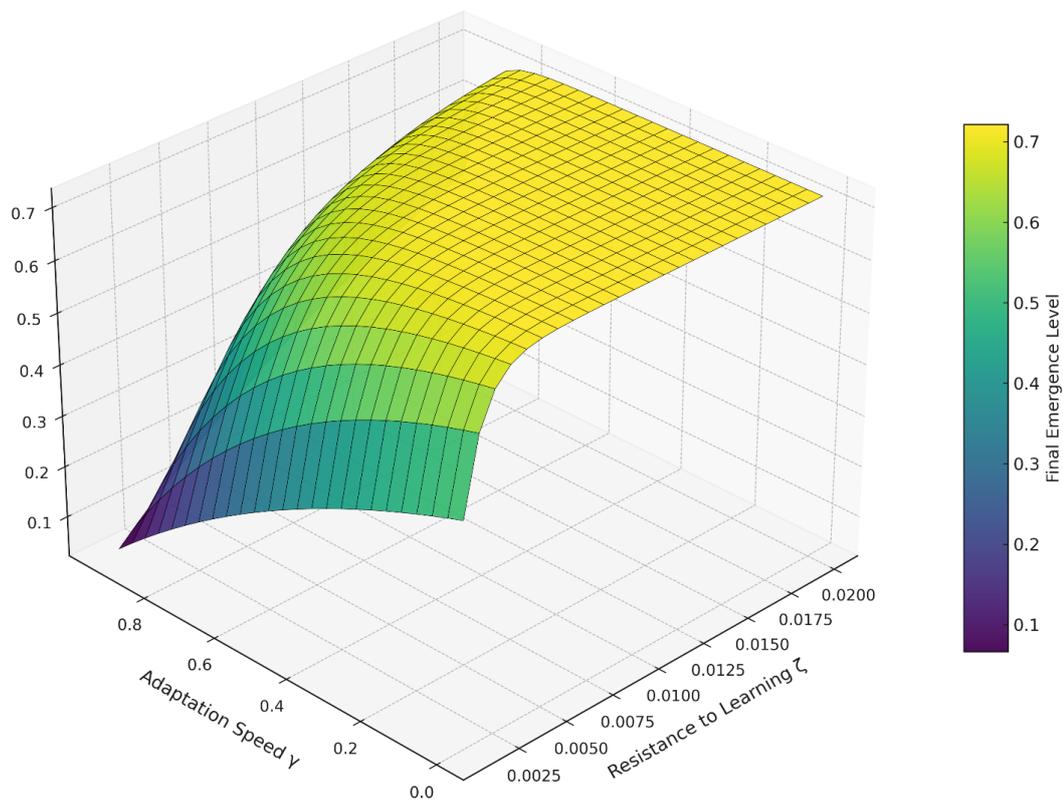

The combination of parameters determines the shape of the phase transition: an S-shaped curve emerges under balanced conditions,

A flat curve appears under high resistance and weak information,

A sharp curve results from strong adaptation and directional external influence.

Notably, randomness does not disappear on its own. It is filtered and collapsed under the influence of adaptation, external fields, and coordination. The phase transition, therefore, is not a single-factor event but rather a multidimensional process that is explicitly dependent on the nature and strength of the interactions between parameters. Depending on their configuration, the system can either remain trapped in chaos or rapidly self-organize.

**5. Discussion**

The phase transition from chaotic ant wandering to coordinated action gradually occurs. However, within this transition lies a range of emergence. Initially, the beetle's movement—caused by many ants—is chaotic (multidirectional), but over time, ant behavior becomes more organized, and the beetle's movement becomes more directed. Eventually, the ants (with few exceptions) concentrate on one side and collectively push the beetle in a single, correct direction—toward the anthill along a trail.

Thus, the emergence range is the critical region during the phase transition where the system reorganizes. First, the beetle moves randomly because it is pulled in different directions. At a certain point, this chaotic influence weakens, and a dominant movement vector begins to form. Finally, nearly all the ants gather on one side, ensuring almost unidirectional motion of the beetle.

We define the emergence range as a transition phase in which the ant–beetle system "decides" which direction will dominate (Table 9).

*Table 9.* Emergence Range as the Region Between Chaos and Stable Order

| | | |
|---|---|---|
| Chaotic phase ($N < 2000$) | Ants act randomly, the directions of force are not coordinated | The beetle moves chaotically, its trajectory is disordered |
| Emergence range ($2000 < N < 7000$) | Filtering of random directions begins, a common vector of movement is formed | The beetle stops moving chaotically, its trajectory becomes more directed |
| Order phase ($N > 7000$) | The ants are organized, the collective movement is stable | The beetle confidently moves towards the anthill |

Emergence can be quantitatively interpreted through movement fluctuations of the beetle:

$$\sigma^2 = \frac{1}{T} \sum_{t=1}^{T} (x_t + x)^2$$

where
σ2 - variance, measuring the degree of chaotic movement;
$x_t$ - beetle position at time ttt;
x¯ - average position over a given time period.

Here, the variance reflects the entropy level in the ant–beetle system. In the chaotic phase, the variance is maximal (random movement). In the emergence range, the variance decreases but remains relatively high. In the ordered phase, the variance tends to zero, as movement becomes almost strictly unidirectional. Therefore, the emergence range corresponds to the moment when the movement variance begins to decrease sharply (Figure 9).

*Figure 9.* Emergence Range as the Transition from Chaotic to Ordered Movement

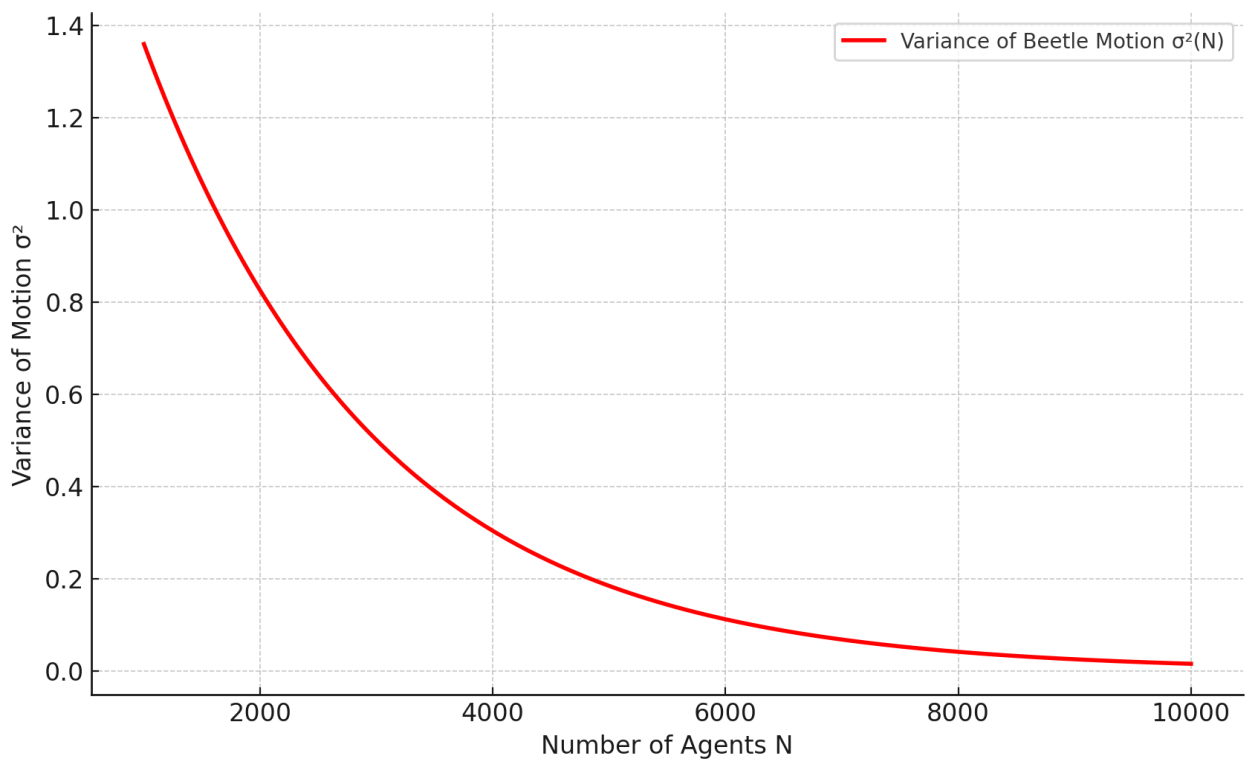

The sharp drop in variance marks the point at which chaotic motion transitions into structured, directed motion.

This is the core principle of vector dissipation of randomness (VDR) as a process through which many random behavioral vectors, inherent to individual agents in a complex system, collapse into a stable and meaningful structure via mechanisms of individual learning, resonant copying, and weak-signal filtering.
In other words, systems composed of many independent agents can generate global behavioral algorithms without centralized control. This is the essence of emergence and self-organization in collective systems.

VDR is thus based on three fundamental mechanisms:

*Individual stochastic learning*
*resonant imitation*
*Weak signal filtering*

When the number of agents N reaches a critical threshold $N_c$, the system transitions from randomness to structured behavior, collapsing into a unified strategy.

In our interpretation, vector dissipation of randomness occurs in twelve stages (Table 10).

*Table 10.* Twelve Stages of Vector Dissipation of Randomness

| Stage | Microlevel process (individual agent) | Process at the meso level (group of agents) | Process at the macro level (the whole system) |
|---|---|---|---|
| 1. Pure chance (Initiation of chaos) | Each agent acts randomly, choosing a vector from sets possible actions | A group of agents exhibits chaotic behavior, without correlations | The entire system is in a maximally entropic state, there are no predictable patterns |
| 2. Local trial of random vectors | The agent tries random vectors, explores the environment, but without feedback yet | The group of agents begins to show the first local variations, but there is no stable trend | Random fluctuations appear in the system, but their amplitude is small |
| 3. First positive signals (Local successes) | Some vectors give an improvement, the agent begins to remember them | The first nodes of local coordination appear - small groups of agents begin to use similar strategies | The system begins to develop nascent structures, but they are still unstable |
| 4. Primary Resonance Points (Local Coordination) | The agent feels that its actions coincide with the successful actions of its neighbors | The group begins to strengthen positive vectors, but they still compete | Local coordination centers appear in the system, but global chaos remains |
| 5. Signal amplification (Group resonance) | The agent focuses on the amplified signals, forming an early algorithm | Correlated behavior begins to form in the group | The system is showing stable patterns for the first time, but they are still unstable |
| 6. Mass filtering of weak vectors (First collapse of randomness) | Vectors that do not lead to success begin to weaken and disappear | The number of random trajectories in the group is significantly reduced, and stable patterns appear | The entropy level in the system decreases and directed movement appears |
| 7. Collective choice (Collapse of randomness) | Agents independently come to similar conclusions that one vector is more effective than others | Almost all agents act the same, with minor fluctuations | The system is "frozen" in one stable pattern of behavior |
| 8. Consolidation of the structure (Collective stabilization) | Every agent makes the right decisions faster | In a group, the algorithm becomes fully ingrained, and random trajectories disappear | The system is completely self-organized, there is no more chaos |
| 9. Phase adaptability (Flexibility to the environment) | The agent responds to changes by adapting its behavior, but within the overall structure | The group may change strategy if new signals emerge | The system demonstrates adaptive stability, but does not break the old algorithm |
| 10. Resonant regeneration (Switching algorithms) | If the environment changes abruptly, the agent is able to return to searching for a new algorithm | The group can create new behavior vectors if the old algorithm becomes ineffective | The system goes into adaptive mode, switching between the VDS phases |
| 11. Long-term self-organization (Emergent optimization) | Agents maintain structure over time but allow microchanges | The group undergoes minimal variations of the algorithm, maintaining its relevance | The system maintains dynamic equilibrium without becoming rigid |

| 12. Collective Intelligence (Global Emergence) | Agents not only optimize their behavior but also form a new cognitive network | The group exhibits hybrid behavior, combining local and global learning | The system functions as a single whole, maintaining adaptability and stability |

This processing can be illustrated as follows (Figure 10):

*Figure 10*. Vector dissipation of randomness culminating in an emergent rational state in the ant–beetle system

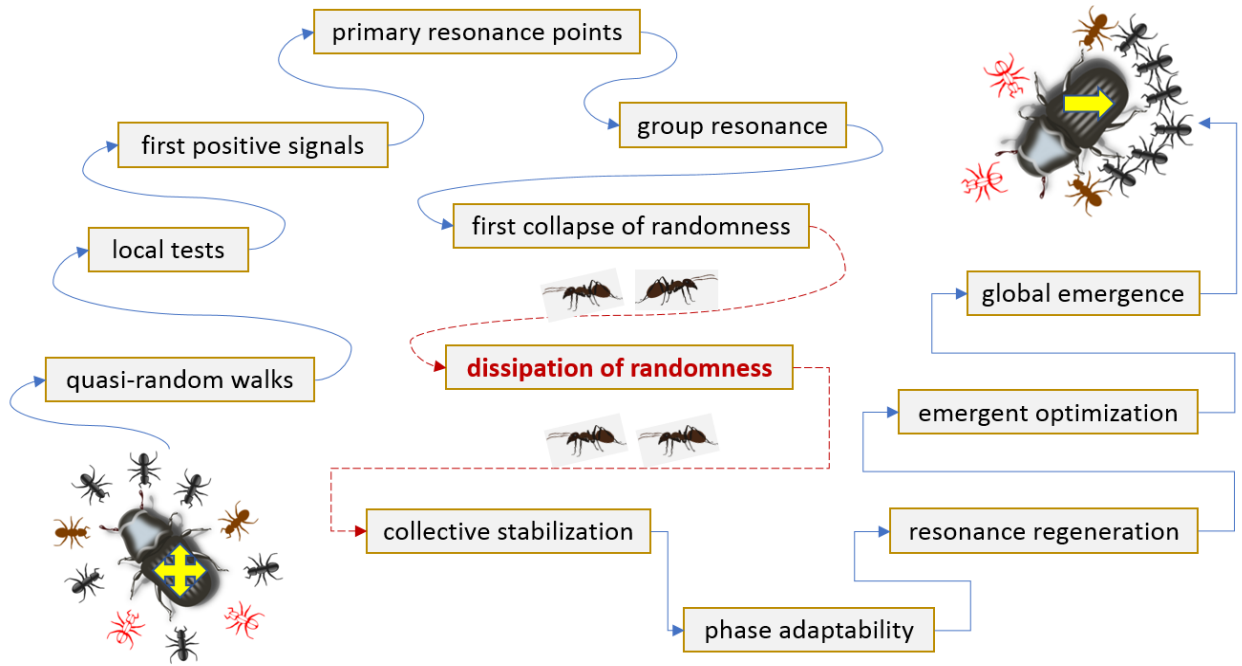

We interpret this emergent transformation in the ant–beetle system as the appearance of what we call paraintelligence.

We define paraintelligence as a functional emergent state that is similar to or in some cases equivalent to—rational activity but without reflexive consciousness.

This raises a fundamental question: does the manifestation of emergence in the ant–beetle system imply the emergence of ant paraintelligence in some form of rational-like functionality?

A robot that delivers an object to a designated location does not possess paraintelligence in this sense since its rational functionality is programmed by human intelligence. In contrast, the ant–beetle system operates without centralized control or external intelligent algorithms.

In our context, emergence refers to the spontaneous development of intentionality, goal-directedness, coordination, learning, adaptation, and environmental influence in an agent-based system. This sets the conditions for what is often called collective intelligence—when a group can act rationally in ways that individual members cannot.

What we observe here is distributed cognition, interpreted through the distribution of functional roles among the group's members.

All of these nonlinear interactions give rise to a form of rationality, but not consciousness. Ants, although they act "rationally," do not possess self-awareness. They have no reflexive understanding of cause–and–effect relationships in their actions. They don't know they are doing the "right" thing. They do not create narratives. They do not possess a map. Neither the individual ant nor the colony has a conscious goal of transporting the beetle to the nest. Their behavior remains chaotic, and goal formation never actually occurs throughout the entire process—from finding the beetle to delivering it to the nest.

Goal-directedness and purpose, therefore, emerge as properties of the collective structural entity, not as intentional acts of individuals.

This is what we call paraintelligence—a form of functional cognition without the overlays or options of subjectivity, reflexivity, consciousness, or self-awareness. It is the result of vector dissipation of randomness (VDR), where numerous local stochastic interactions collapse into a coherent, directed vector of collective agent behavior.

It follows logically that intelligence does not necessarily require consciousness, and consciousness is not a prerequisite for intelligence. Intelligence can exist without reflexive awareness or self-consciousness. It may be emergent, vectorial, distributed, subconscious, nonsubjective, or purely functional.

In its simplest form, paraintelligence can be expressed as a modular cognitive architecture composed of six fundamental blocks (Figure 11):

1. *Random Micro-Acts – agents act independently and chaotically*
2. *Local Amplification – positive random actions are locally reinforced*
3. *Suppression of noise – ineffective vectors are gradually suppressed*
4. *Emergence of Pattern – a structure begins to emerge at the macro level*
5. *Directional Consensus – a majority of agents act in coordinated ways*
6. *Emergent goal behavior—the agent collective achieves a meaningful outcome without knowing the goal*

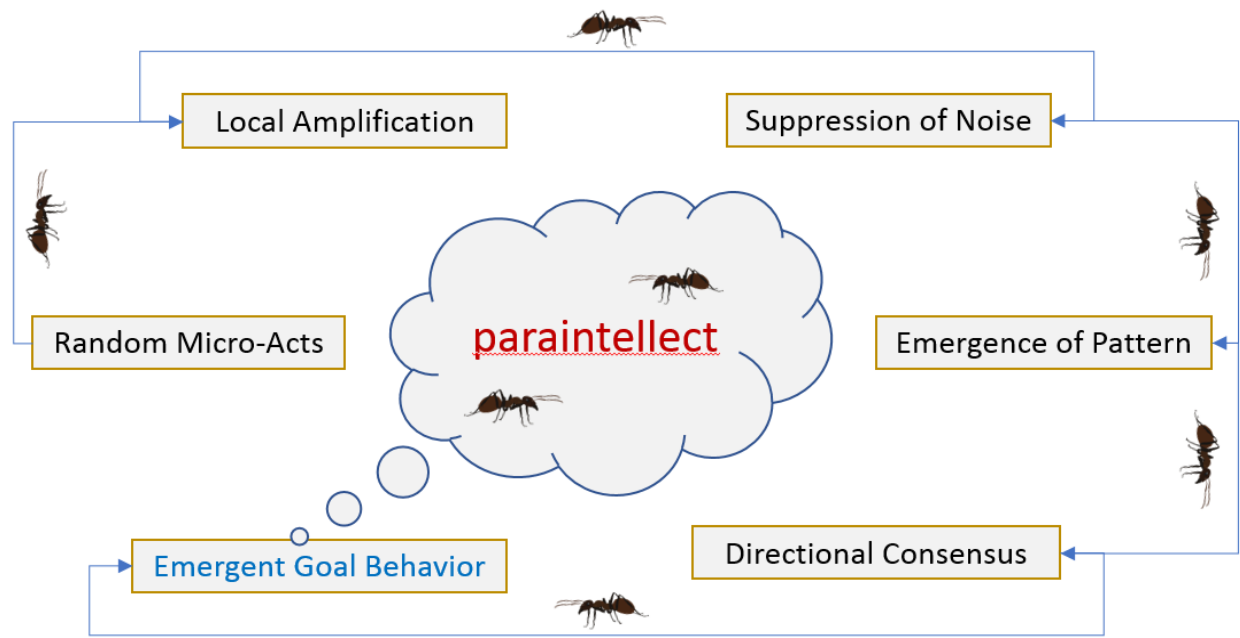

Figure 11. Architecture of Paraintelligence

In this sense, paraintelligence is a *function of form*, without a *function of content*—a structural intelligence without consciousness that exists solely through situational and transient structured interactions.

It arises in behavioral systems when disorder flows into order, dispersion transforms into a structure, and reactivity evolves into directedness and implicit goal formation. In this context, we may interpret paraintelligence as a type of cognitive architecture in which:

*thinking is replaced by correlation,*
*goals are replaced by outcomes,*
*and will be replaced by the density of accumulated environmental stimuli.*

We can argue that such decentralized and nonanthropomorphic paraintelligence is a common and natural phenomenon in biological systems.

## 6. Conclusion

In this study, we examined the mechanism of vector dissipation of randomness (VDR), which interprets the transition of the complex *ant–beetle* system from chaotic behavior to organized collective action through dynamic filtering of random strategies and the accumulation of information in the environment.

In the system, individual random strategies of agents undergo selection: ineffective movement directions are filtered out, whereas successful ones are reinforced. Environmental information accumulates through feedback (via pheromone trails and interactions with neighboring agents). The system undergoes a gradual phase transition, which is not instantaneous but represents a cumulative process of ordering. The critical number of agents determines the point at which the system transitions to a coordinated state. The rate of randomness filtration depends on the agents' adaptability and connectivity.

The newly proposed normalized emergence function (NEF) interprets emergence through three core dependencies: directional correlation among agents, environmental information accumulation, and adaptive transformation of agent behavior.

Numerical experiments confirmed the presence of a smooth phase transition in the VDR, with a critical agent number for stable self-organization of approximately $N \approx 5000$. They also revealed an exponential increase in the beetle's movement speed once this critical threshold was surpassed.

VDR differs significantly from traditional models of phase transition (e.g., the Ising model, swarm intelligence, and percolation). Unlike these models, which typically involve abrupt or discrete transitions, the VDR demonstrates a progressive compression of random strategies into a single dominant vector.

In classical models (such as the Ising model or percolation), a new emergent state appears sharply or suddenly. In contrast, in the *ant–beetle* system, no such sudden jump in coordination occurs. Instead, emergence is present from the very beginning—as one of many random behavioral vectors that gradually becomes dominant as the others dissipate.

In simpler terms, emergence is always present—hypothetically or functionally. The question is under what conditions and when it becomes perceptible or statistically significant.

In classical swarm intelligence models (e.g., particle swarm optimization (PSO)), agents operate within rigid rules but without explicit interpretation of the phase transition (Table 11). In ant colony optimization (ACO) models, pheromones play a dominant role, but agent evolution over time is not explicitly addressed.

*Table 11.* Comparative Interpretation of Self-Organization and Emergence Mechanisms

| Model | How is self-organization achieved? | How is a phase transition described? | Application |
| --- | --- | --- | --- |
| VDR (Vector Dissipation of Randomness) | Agents filter random directions, adapt, and reinforce correct strategies | Smooth phase transition through emergence | Biology, sociodynamics, AI, ecosystems |
| ACO (Ant Colony Optimization) | Ants leave pheromones, strengthen shortest paths | The phase transition is abrupt: if one path is stronger, the others disappear. | Route optimization, logistics |
| PSO (Particle Swarm Optimization) | Particles update speed based on group leaders | Continuous phase transition, but without critical point | Machine learning, feature optimization |
| SOM (Self-Organizing Maps) | The neurons of the network are grouped into a topological map | Gradual increase in complexity of clustering | Data clustering, image analysis |
| Ising model | The interaction of spins leads to magnetic ordering | Classical abrupt phase transition | Physics, modelling of collective behavior |

Thus, NEF combines and adapts elements from ACO (direction filtering), PSO (agent adaptation), and SOM (multidimensional self-organization).

In ant-based learning, other models, such as reinforcement learning, realize agent adaptation through optimization routines, not through behavioral correlation, as in NEF.

Specifically, self-organizing maps (SOMs) do not account for environmental influence (Table 12), unlike NEF.

*Table 12*. Comparative Interpretation of Agent Motion Control Mechanisms

| Model | How do agents make decisions? | Is there training? | The role of the environment |
|---|---|---|---|
| VDR | Filtering random directions, feedback | Yes (adaptation of strategies) | Influences through accumulation of information (pheromones, density) |
| ACO | Pheromone-based path selection | No (pheromones are the only mechanism) | The environment influences through pheromones |
| PSO | Particles move to the best points | Partially (each particle takes into account the successes of its neighbors) | There is no obvious influence of the environment |
| SOM | Neurons adjust weights based on input data | Yes (self-learning without a teacher) | There is no obvious influence of the environment |
| Ising | Spins change orientation depending on their neighbors | No | The environment influences through the external magnetic field |

Accordingly, the NEF within the VDR framework is a model where agents learn, change strategies, and respond to environmental influence.

We can highlight the following dominant features and paradigms of the normalized emergence function:

unlike critical models (e.g., ACO, Ising), NEF produces a smooth phase transition, not a discrete jump;

Unlike PSO, NEF accounts for environmental factors (pheromones, agent density);

SOM and PSO lack an explicit phase transition, whereas NEF includes a clearly defined emergence range;

NEF is the only model in which agents learn, adjust strategies, and respond to the environment.

Overall, we conclude that NEF more authentically and adequately models biological and sociodynamic systems, which typically lack instantaneous transitions. Instead, order emerges from chaos through gradual information accumulation. This applies even to so-called "instantaneous" biological mutations. In this sense, the authenticity and adequacy of NEF extends to all systems with gradual adaptation, including biology, artificial intelligence, cognitive models, and complex networks.

For paraintelligence, it represents a new emergent state arising in complex, nonlinear systems with rational outcomes. This state is not limited to biological systems—it also characterizes synthetic systems, including those based on artificial intelligence.

We may define artificial intelligence (AI) as a decentralized and algorithmic carrier of intelligence without consciousness or self-awareness that is capable of functionally rational behavior in fixed or open environments. This also applies to many nonanthropomorphic biological systems.

Therefore, AI and biological paraintelligence do not need to be human-like to be recognized as intelligent. They are already functionally intelligent if their behavior demonstrates persistent directionality and adaptiveness. After all, consciousness is not a prerequisite for intelligence; rather, consciousness and self-awareness are among its many possible modes or options.